\documentclass[12pt,preprint]{aastex}
\def\degpoint{\ifmmode ^{\rm{o}}\!. \else $^{\rm{o}}\!.$\fi}
\newcommand{\degrees}{$^{\rm{o}}$}
\newcommand{\ms}{\mbox{m\ s$^{-1}$}}
\newcommand{\kms}{\mbox{km \ s$^{-1}$}}
\newcommand{\Msun}{\mbox{M$_{\odot}$}}

\newcommand{\Mjup}{\mbox{M$_{\rm Jup}$}}
\newcommand{\Rjup}{\mbox{R$_{\rm Jup}$}}

\newcommand{\feh}{\mathrm{[Fe/H]}}
\newcommand{\teff}{T_\mathrm{eff}}
\newcommand{\logg}{\log g}
\newcommand{\fei}{Fe~\textsc{i}}
\newcommand{\feii}{Fe~\textsc{ii}}
\newcommand{\hst}{\textit{Hubble Space Telescope} }
\newcommand{\ltsimeq}{\raisebox{-0.6ex}{$\,\stackrel
         {\raisebox{-.2ex}{$\textstyle <$}}{\sim}\,$}}
\newcommand{\gtsimeq}{\raisebox{-0.6ex}{$\,\stackrel
         {\raisebox{-.2ex}{$\textstyle >$}}{\sim}\,$}}

\begin{document}
\title{HD 91669b: A New Brown Dwarf Candidate from the McDonald 
Observatory Planet Search}
\author{Robert A.~Wittenmyer\altaffilmark{1,2}, Michael 
Endl\altaffilmark{1}, William D.~Cochran\altaffilmark{1}, Ivan 
Ram\'irez\altaffilmark{1,3}, Sabine Reffert\altaffilmark{4}, Phillip 
J.~MacQueen\altaffilmark{1}, Matthew Shetrone\altaffilmark{1} }
\altaffiltext{1}{McDonald Observatory, University of Texas at Austin, 
Austin, TX 78712}
\altaffiltext{2}{Department of Astrophysics, School of Physics, 
University of NSW, 2052, Australia}
\altaffiltext{3}{Max Planck Institute for Astrophysics, Postfach 1317, 
Garching, Germany}
\altaffiltext{4}{Zentrum f{\" u}r Astronomie Heidelberg, Landessternwarte, 
K{\" o}nigstuhl 12, 69117 Heidelberg, Germany}
\email{rob@phys.unsw.edu.au}

\shorttitle{Brown Dwarf Orbiting HD 91669}
\shortauthors{Wittenmyer et al.}
\begin{abstract}

\noindent We report the detection of a candidate brown dwarf orbiting 
the metal-rich K dwarf HD~91669, based on radial-velocity data from the 
McDonald Observatory Planet Search.  HD~91669b is a substellar object in 
an eccentric orbit ($e$=0.45) at a separation of 1.2~AU.  The minimum 
mass of 30.6~\Mjup\ places this object firmly within the brown dwarf 
desert for inclinations $i\gtsimeq$23\degrees.  This is the second rare 
close-in brown dwarf candidate discovered by the McDonald planet search 
program.

\end{abstract}


\section{Introduction}

Brown dwarfs are commonly defined as substellar objects with masses 
between the deuterium-burning limit (about 13\Mjup) and the 
hydrogen-burning limit (about 80\Mjup).  Despite the large (several 
hundred \ms) reflex radial-velocity signals produced by brown dwarfs in 
orbit around main-sequence stars, planet-search surveys monitoring 
thousands of stars have found remarkably few such objects in orbits with 
$a\ltsimeq$5~AU.  By contrast, nearly 300 planets have been found by the 
radial-velocity method.  Early radial-velocity surveys by 
\citet{campbell88} and \citet{murdoch93} reported a distinct lack of 
candidate companions in the brown dwarf mass range, despite their 
relatively easy detectability.  \citet{campbell88} noted that 7 of 15 
targets showed velocity variability consistent with a distant companion, 
but the lack of astrometric variability led those authors to limit the 
companion mass range to $\sim$1-9 \Mjup.  In the nearly 20 years spanned 
by current radial-velocity planet searches, only a handful of brown 
dwarf candidates have been identified, e.g.~HD~114762b (11 \Mjup; Latham 
et al.~1989, Cochran et al.~1991, Mazeh et al.~1996), HD~168443c (18 
\Mjup; Marcy et al.~2001), HD~202206b (17 \Mjup; Udry et al.~2002), and 
HD~137510b (26 \Mjup; Endl et al.~2004).  Recently, a transiting brown 
dwarf has been discovered by the \textit{CoRoT} spacecraft.  This 
object, CoRoT-Exo-3b, has a mass of 21.7 \Mjup\ and a radius of 1.01 
\Rjup, in a remarkably close orbit at 0.05~AU \citep{deleuil08}.

The steep rise in the planetary mass function toward lower masses 
\citep{marcy05a} combined with the decline in lower-mass stellar 
companions \citep{mazeh03} give evidence for the existence of a ``brown 
dwarf desert'': a paucity of brown dwarf companions orbiting within 
$\sim$3-4~AU of solar-type stars \citep{mb00}.  In contrast, there is no 
dearth of brown dwarfs orbiting at larger separations \citep{gizis01}, 
or free-floating \citep{reid99}.  \citet{grether06} performed a detailed 
investigation of the planetary and stellar mass distributions in order 
to correct for biases and to assess the reality of the brown dwarf 
desert.  Defining close companions as those with orbital periods 
$P<$5~yr, \citet{grether06} confirmed the deficit of close brown dwarf 
companions and estimated the driest part of the ``desert'' to lie at 
$M=31^{+25}_{-18}$ \Mjup.

It is possible that selection biases curtail the discovery rate of brown 
dwarfs relative to planets, as measured by published results.  The push 
to detect lower-mass planets combined with the high value of telescope 
time lead planet search teams to discard targets which exhibit the 
large-amplitude radial-velocity variations induced by brown dwarf 
companions in close orbits.  However, a simple bias of this nature does 
not seem likely to explain the factor of $\sim$100 deficit in the 
observed frequency of brown dwarf companions compared to planetary 
companions.

Since the radial-velocity method yields only a minimum mass for the 
companion, some number of brown dwarf candidates are likely to be 
low-mass stars orbiting at low inclinations.  Some authors have combined 
the radial-velocity orbits with astrometric data in order to determine 
the inclination and true mass of the companions.  In this way, the 
``planets'' HD~38529c and HD~168443c were revealed to be brown dwarfs 
with masses of 37 \Mjup\ and 34~\Mjup, respectively \citep{reffert06}.  
\citet{bean07} used \hst astrometry to determine that the planet 
candidate HD~33636b, with a minimum mass of 9.3~\Mjup, was in fact a 
star with a true mass of 142 \Mjup\ orbiting at a nearly pole-on 
inclination $i=4.1$\degr.  These examples illustrate the importance of 
finding brown dwarf candidates with low values of m~sin~$i$ to increase 
the likelihood of their true mass remaining in the substellar mass 
range.

In this paper, we derive the parameters of the host star HD~91669 
(\S~3), and in \S~4 we present evidence for its brown dwarf companion, 
and discuss possible constraints on its true mass.

\section{Observations and Data Analysis}

Observations were made from the 2.7m Harlan J.~Smith Telescope at 
McDonald Observatory using the 2dcoud\'e echelle spectrograph 
\citep{tull94} at a resolution $R=\lambda/\Delta\lambda=60,000$.  
HD~91669 is one of $\sim$300 stars being monitored in the long-term 
planet search program, which began in 1988.  Observations with the 
current instrumental configuration (``Phase~III'') began in 1998 and 
achieve a routine internal precision of 6--9 \ms.  Complete details of 
the planet search at McDonald Observatory are given in 
\citet{limitspaper}.  HD~91669 was added to the program in 2004 March 
after a preliminary abundance analysis showed it to be a metal-rich 
star.  At $V=9.7$, exposure times ranged from 20-30~minutes per epoch.  
Radial velocities were obtained using a temperature-stabilized iodine 
($I_2$) cell, imprinting thousands of narrow absorption lines on the 
stellar spectrum and providing a velocity metric \citep{butler96}.  The 
iodine lines were also used to model the effects of the spectrograph's 
instrumental profile, as described by \citet{valenti95} and 
\citet{endl00}.  We fit Keplerian orbits to the radial-velocity data 
using \textit{Gaussfit}, a least-squares and robust-estimation package 
\citep{jefferys87}.

\section{Stellar Parameters}

The stellar parameters $\teff$, $\logg$, and $\feh$ (Table~1) were 
determined using a standard iterative spectroscopic procedure, as 
described in Ram\'irez et al. (2007). The transition probabilities 
adopted for the iron lines were taken from laboratory measurements; 
i.e., no astrophysical $\log gf$ values were used. Effective 
temperatures were obtained from as many as possible of the 
color-temperature relations by Ram\'irez \& Mel\'endez (2005).  Surface 
gravities were determined from the location of the stars on the HR 
diagram, on which the theoretical isochrones of Bertelli et al.~(1994) 
were superimposed. We used the accurate \textit{Hipparcos} parallaxes of 
the stars to calculate their absolute magnitudes before determining 
$\logg$.  In a similar manner, the stellar masses and radii (Table~1) 
were found using the derived stellar parameters and the Bertelli et 
al.~isochrones.  Details on the adopted method of $\logg$, mass, and 
radius determination are given by Reddy et al.~(2003) and Allende Prieto 
et al.~(2004).

The iron abundance was inferred from a large set of unblended \fei\ 
lines covering a wide range of line strength and about a dozen \feii\ 
lines, which allowed us to determine accurate microturbulent velocities 
($v_t$) and check the ionization equilibrium of iron lines, 
respectively.  After applying the empirical correction to the iron 
abundance determined from \fei\ lines suggested by Ram\'irez et al. 
(2007), the mean iron abundances determined separately from \fei\ and 
\feii\ lines are typically in agreement and their average value is 
adopted as the star's $\feh$.  For HD~91669, however, the mean \feii\ 
abundance was about 0.3 dex larger than the mean \fei\ abundance, a 
systematic error occurring in cool metal-rich dwarf stars which has been 
reported by several authors (e.g., Yong et al.~2004, Allende Prieto et 
al.~2004, Schuler et al.~2006, Ram\'irez et al.~2007) and has been 
suggested to be due to non-LTE effects or inadequacies in the modeling 
of cool stellar atmospheres.  Given that \fei\ is the dominant species 
at these cool temperatures, the mean iron abundance from \fei\ lines 
only was adopted as the star's [Fe/H].

HD~91669 (=HIP~51789, NLTT~24745) is a metal-rich ([Fe/H]=+0.31) K0 
dwarf which has been in the planet search program at McDonald 
Observatory for 4.3~yr.  It has a \textit{Hipparcos} parallax of 
11.58$\pm$1.48 mas, corresponding to a distance of 86.4 pc 
\citep{newhipp}.  The derived $\logg$ of 4.48 refutes the K0/1~III 
classification reported by \citet{houk88}.  The uncertainties on $\teff$ 
and $\logg$ given in Table~1 represent internal errors, and do not 
include possible systematic errors of $\sim$100~K in $\teff$ and 
$\sim$0.1~dex in $\logg$.

\section{Companion Parameters}

We present 18 radial-velocity observations of HD~91669 spanning 4.3~yr 
which indicate a massive substellar companion (Figure~1).  For the 
adopted stellar mass $M_*=0.914$\Msun, HD~91669b has a minimum mass of 
m~sin~$i=30.6\pm2.1$\Mjup\ (Table~2).  With a separation $a$=1.205~AU, 
and a distance of 86.4 pc, the angular separation of HD~91669b would be 
14 mas.  It is conceivable that a stellar companion may be ruled out by 
direct imaging, but the 14~mas projected separation of HD~91669b is 
considerably smaller than the limits achieved by current AO and 
coronagraphic surveys (e.g.~Nielsen et al.~2008).
For randomly distributed inclinations, the mean value of sin~$i$ is 
$\pi/4=0.785$.  Applying this adjustment to the minimum mass, the true 
mass of HD~91669b would be 39.0\Mjup, remaining comfortably below the 
hydrogen-burning limit of 80\Mjup.  For a true mass in the stellar 
regime, the inclination must be $i<22.5$\degrees.  The probability that 
such an object would have that inclination or smaller is given by
\begin{equation}
\textrm{Prob}(i<i_c)=1-\cos(i).
\end{equation}
For HD~91669, this \textit{a priori} probability is 7.6\%.  
\citet{kurster08} were able to combine \textit{Hipparcos} astrometry 
with a radial-velocity orbital solution to place additional constraints 
on the inclination of GJ~1046b, a brown dwarf candidate in close orbit 
about an M2.5 star.

Similarly, we also tried to fit the Hipparcos Intermediate Astrometric 
Data based on the new reduction of the Hipparcos raw data by van Leeuwen 
(2007) for HD~91669 (HIP~51789) with an orbital model. We fixed the five 
orbital parameters derived from the radial velocities (Table~2), and 
fitted for the two missing orbital parameters (inclination and ascending 
node), and simultaneously for corrections to the five standard 
astrometric parameters (positions, proper motions, parallax), in an 
effort to get a better constraint on especially the inclination and thus 
on the companion mass. Note that in order to derive constraints on the 
inclination, the orbit does not need to be detected; often, small 
inclinations (corresponding to large companion masses) can be excluded 
(see Reffert \& Quirrenbach 2009) because they would correspond to 
astrometric signals much larger than the measurement precision of 
Hipparcos, which would be seen in the data. For HD~91669, inclinations 
smaller than 5.3\degr\ and larger than 176.2\degr\ can be excluded with 
99.73\% confidence, corresponding to a companion mass smaller than about 
0.6~\Msun.  This is not a good constraint on the companion mass, 
but it is all that we can safely derive from the Hipparcos data. There 
are 81 field transits (individual 1-dimensional measurements) for 
HD~91669 with a median formal error of 6.0~mas. In contrast to that, the 
minimum astrometric signature of the companion, assuming an inclination 
of 90\degr, is only 0.9~mas peak-to-peak. It is clear that this would be 
too small to be significantly detectable by Hipparcos. Thus, while we 
cannot derive a tight constraint on the companion mass, the Hipparcos 
data are at least not inconsistent with a companion mass in the brown 
dwarf regime.

The orbital solution has an rms of 6.3 \ms, consistent with our typical 
velocity precision for fainter stars in the 2.7m planet search program 
(HD~91669: $V$=9.70), and considerably smaller than the mean uncertainty 
of 10.0$\pm$1.2 \ms.  The uncertainties on the Keplerian orbital 
parameters are derived in two ways: 1) \textit{Gaussfit} returns 
uncertainties generated from a maximum likelihood estimation that is an 
approximation to a Bayesian maximum a posteriori estimator with a flat 
prior \citep{jefferys90}, and 2) we applied a Monte Carlo method similar to 
that of \citet{marcy05b}.  We determine the best-fit set of model 
parameters and subtract the Keplerian model from the data, generating a 
residuals file.  Then we create simulated data sets retaining the epochs 
of observation, where the velocity at each point consists of the 
best-fit Keplerian model added to a residual velocity randomly drawn 
(with replacement) from the residuals file.  In this way, each simulated 
velocity point is ``bumped'' by an amount consistent with the variations 
due to stellar jitter and instrumental errors.  We generate 100 such 
realizations, fit each with a Keplerian model as above, and record the 
derived parameters.  The 1-$\sigma$ uncertainty of each parameter is 
then taken to be the standard deviation about the mean value of the 100 
sets of parameters.  The parameter uncertainties shown in Table~2 are 
those obtained from the latter method.  We note that the uncertainties 
of orbital parameters derived from least-squares fitting are known to be 
non-Gaussian \citep{ford05}, and hence can be underestimated by a factor 
of 5-10 \citep{otoole08}.  The radial-velocity data are given in 
Table~3.  The residuals to the orbit fit show no slope or evidence of 
residual signals; the highest periodogram peak has a false-alarm 
probability (FAP) of 95\%.

\section{Summary}

We have presented radial-velocity observations indicating the presence 
of a rare brown dwarf candidate orbiting the metal-rich K dwarf HD~91669.  
HD~91669b is the second brown dwarf identified in the McDonald 
Observatory planet search program.  Like its predecessor HD~137510b 
\citep{endl04}, this object has a relatively low mass and hence a high 
probability of having a true mass within the substellar regime.  We note 
that the rarity of brown dwarfs implies that the detection of two 
candidates in a sample of 250 stars is rather unlikely.  However, the 
McDonald Observatory program yields a detection rate of 0.8$\pm$0.6\%, 
compared to a rate of 0.7$\pm$0.2\% from the California \& Carnegie 
Planet Search (7 candidates from $\sim$1000 target stars: Vogt et 
al.~2002, Patel et al.~2007).  Within the limits of small-number 
stattics, the yields of brown dwarf candidates from these two programs 
are comparable.

\acknowledgements

This material is based on work supported by the National Aeronautics and 
Space Administration under Grants NNG04G141G, NNG05G107G issued through 
the Terrestrial Planet Finder Foundation Science program and Grant 
NNX07AL70G issued through the Origins of Solar Systems Program.  We are 
grateful to the McDonald Observatory TAC for their generous allocation 
of telescope time for this project.  This research has made use of 
NASA's Astrophysics Data System (ADS), and the SIMBAD database, operated 
at CDS, Strasbourg, France.



\begin{deluxetable}{ll}
\tabletypesize{\scriptsize}
\tablecolumns{2}
\tablewidth{0pt}
\tablecaption{HD 91669 Stellar Parameters\label{tbl-1}}
\tablehead{
\colhead{Parameter} & \colhead{Value }  
}
\startdata
T$_{eff}$ & 5185$\pm$87 K \\
log $g$ &  4.48$\pm$0.20 \\
v$_t$ & 0.85 \kms \\
Mass & 0.914$^{+0.018}_{-0.087}$ \Msun \\
$[Fe/H]$ & $+0.31\pm$0.08 \\
log~$R'_{HK}$ & $-4.66\pm$0.18 \\
\enddata
\end{deluxetable}

\begin{deluxetable}{ll}
\tabletypesize{\scriptsize}
\tablecolumns{2}
\tablewidth{0pt}
\tablecaption{HD 91669 Companion Parameters \label{tbl-2}}
\tablehead{
\colhead{Parameter} & \colhead{Value } 
}
\startdata
Period & 497.5$\pm$0.6 days \\ 
$T_0$ & 53298.4$\pm$0.6 JD-2400000 \\
$e$ & 0.448$\pm$0.002 \\
$\omega$ & 161.3$\pm$0.5 \degrees \\
$K$ & 933.0$\pm$3.1 \ms \\
M sin $i$ & 30.6$\pm$2.1 \Mjup \\
$a$ & 1.205$\pm$0.039 AU \\
$\chi^2_{\nu}$ & 0.41 \\
RMS (\ms) & 6.3 \\
$N_{obs}$ & 18 \\
\enddata
\end{deluxetable}

\begin{figure}
\plotone{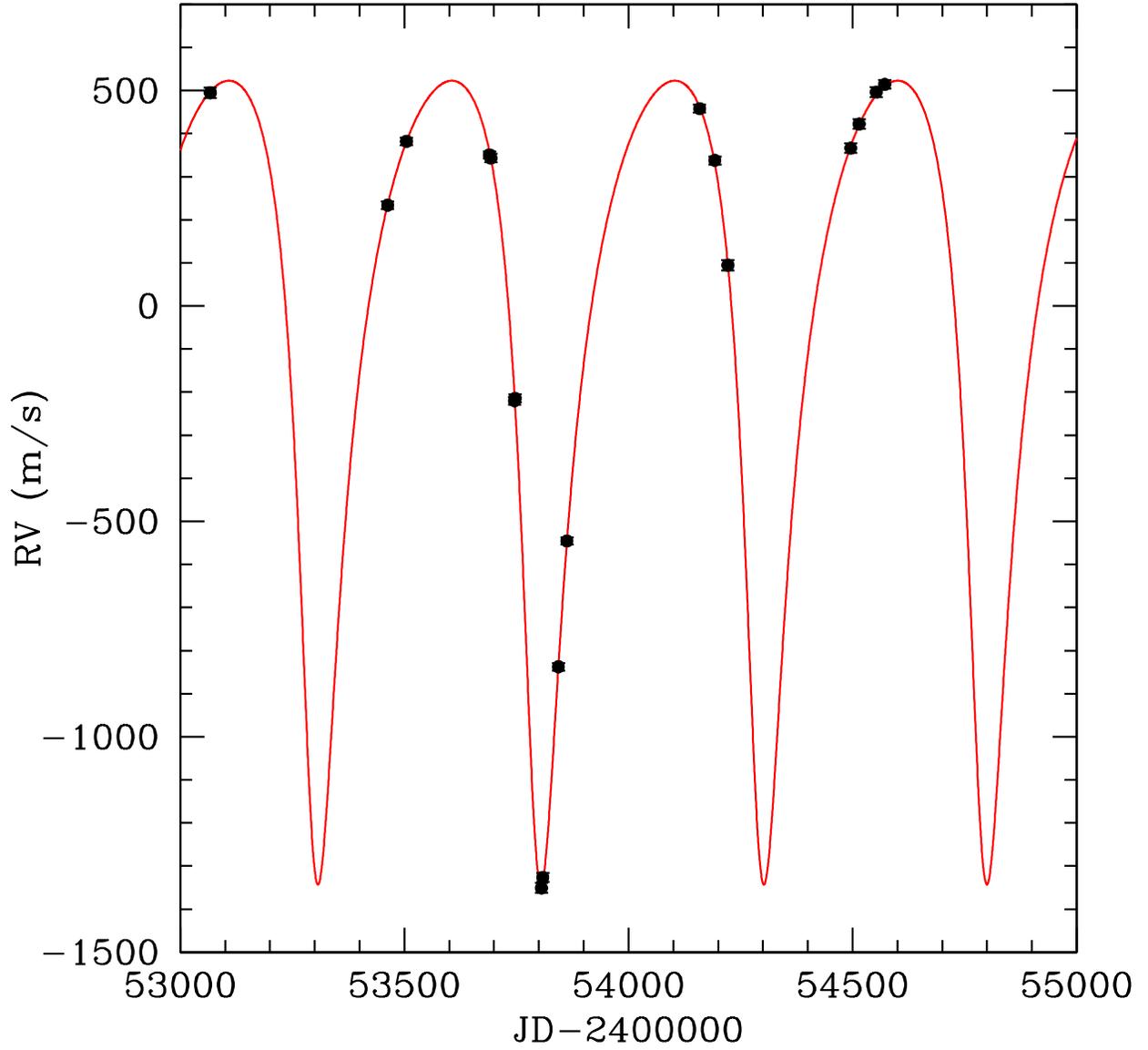}
\caption{Keplerian orbital solution for HD~91669.  The residual rms about 
this fit is 6.3 \ms. } 
\end{figure}

\begin{deluxetable}{lrr}
\tabletypesize{\scriptsize}
\tablecolumns{3}
\tablewidth{0pt}
\tablecaption{2.7m Radial Velocities for HD 91669}
\label{tbl-3}
\tablehead{
\colhead{JD-2400000} & \colhead{Velocity (\ms)} & \colhead{Uncertainty
(\ms)}}
\startdata
53066.87750  &    495.1  &   12.3  \\
53462.78781  &    233.9  &    9.2  \\
53504.63667  &    382.0  &    8.5  \\
53689.98969  &    350.7  &    8.6  \\
53692.99313  &    343.1  &    9.7  \\
53745.96115  &   -220.3  &    9.2  \\
53746.97586  &   -214.2  &    9.2  \\
53805.84071  &  -1351.1  &   11.5  \\
53808.79440  &  -1326.5  &   10.3  \\
53843.70848  &   -837.7  &    8.9  \\
53862.65416  &   -545.6  &    8.4  \\
54158.85204  &    458.1  &    9.5  \\
54192.73118  &    338.0  &    9.3  \\
54221.74547  &     94.8  &   11.8  \\
54495.90692  &    366.4  &   10.6  \\
54514.80082  &    422.5  &   10.9  \\
54552.84010  &    496.4  &   11.5  \\
54571.70919  &    514.4  &    9.9  \\
\enddata
\end{deluxetable}

\end{document}